\documentclass[twocolumn,showpacs,nopreprintnumbers,amsmath,pra,superscriptaddress]{revtex4}
\usepackage{amsmath,amssymb}

\usepackage{graphicx}

\begin{document}

\title{Pure superposition states of atoms generated by a bichromatic
elliptically polarized filed}

\author{A.V. Taichenachev}
\affiliation{Novosibirsk State University, Novosibirsk 630090, Russia} \affiliation{Institute of
Laser Physics SD RAS, Novosibirsk 630090, Russia, e-mail: LLF@LASER.NSC.RU}
\author{V.I. Yudin}
\affiliation{Novosibirsk State University, Novosibirsk 630090, Russia} \affiliation{Institute of
Laser Physics SD RAS, Novosibirsk 630090, Russia, e-mail: LLF@LASER.NSC.RU}
\author{V.L. Velichansky}
\affiliation{P.N. Lebedev Physical Institute RAS, Moscow 117924, Russia}
\author{S.A. Zibrov}
\affiliation{P.N. Lebedev Physical Institute RAS, Moscow 117924, Russia}

\begin{abstract}
We find specific polarizations of components of a bichromatic field, which
allow one to prepare pure superposition states of atoms, using the coherent
population trapping effect. These $m$$-$$m$ states are prepared in the system
of Zeeman substates of the ground-state hyperfine levels with arbitrary angular
momenta $F_1$ and $F_2$. It is established that, in general case $m\ne 0$, the
use of waves with elliptical polarizations
($\varepsilon_1$$\perp$$\varepsilon_2$ field configuration for alkali metal
atoms) is necessary for the pure state preparation. We analytically show an
unique advantage of the D1 line of alkali metal atoms, which consists in the
possibility to generate pure $m$$-$$m$ states even in the absence of spectral
resolution of the excited-state hyperfine levels, contrary to the D2 line.
\end{abstract}


\maketitle

\section{Introduction}
For many applications such as high-precision spectroscopy, frequency standards,
magnetometry, lasers, masers, quantum informatics etc. it is important to have
a possibility to prepare (in a controllable way) pure quantum states of atoms,
which are superpositions of states belonging to different energy levels. The
coherent population trapping (CPT) effect \cite{Alz,Ar1} is one of possible
ways for the superposition state preparation by optical means. In this case the
pure quantum state is a dark (non-absorbing) state. For a model three-state
$\Lambda$ system the CPT effect has been well studied \cite{Ar1,Gr}. However,
for real atoms with complicated Zeeman and hyperfine structures of energy
levels, a nontrivial problem on the choice of excitation scheme suitable for
the pure superposition state generation is actual \cite{St,Van,Taich,Jau,Zan}.

It is worth to formulate two main conditions for the optimal dark resonance detection, when the
resonance amplitude and contrast are maximal. i) At the exact two-photon resonance the pure dark
superposition state has to exist. ii) Trap dark states insensitive to the Raman detuning should
be absent. In opposite case part of atoms will be accumulated in such states in the course of
optical pumping, and they will not contribute to the two-photon resonance formation.

In the present paper we theoretically investigate the possibility of the dark
superposition state preparation in the system of Zeeman substates of the
ground-state hyperfine levels with the total angular momenta $F_1$ and $F_2$.
It is assumed that a bichromatic field excites the two-photon resonance of
$\Lambda$ type. General conditions for the polarizations of the frequency
components are formulated in the case when the pure dark state is a
superposition of the Zeeman substates $|F_1,m\rangle$ and $|F_2,m\rangle$ with
the same magnetic quantum number $m$, i.e. at the condition of the $m$$-$$m$
two-photon resonance. As is well-known, such resonances are used in atomic
clocks and magnetometers based on CPT \cite{Hem,Kit,St2}. The main attention is
given to the case of alkali metal atoms, when the angular momenta $F_1$ and
$F_2$ differ by one. We show that in this case  the $m$$-$$m$ dark states are
generated by the $\varepsilon_1$$\perp$$\varepsilon_2$ configuration of the
bichromatic running wave, where the major semiaxes of the polarization ellipses
are orthogonal. The ellipticity parameters $\varepsilon_{1}$  and
$\varepsilon_{2}$ are coupled by a relationship dependent on the quantum
numbers $m$ and $F$. In the particular case of $m=0$ the pure dark 0$-$0 states
are prepared in the $\varepsilon$$\perp$$\varepsilon$ field configurations,
where the frequency components have the same ellipticity. We consider also the
situation when the excited-state hyperfine structure is not spectrally resolved
(e.g. due to the optical transition broadening by collisions with a buffer
gas). It turns out that for the D1 line of alkali metal atoms all the dark
$m$$-$$m$ states are conserved, while for the D2 line such states are absent.

\section{Statement of the problem}
We consider the resonant interaction of atoms with a two-frequency field formed
by copropagating waves running along $z$ axis:
\begin{equation}\label{1}
{\bf E}(z,t)=E_1(z){\bf a}_1e^{-i\omega_1t}+
E_2(z){\bf a}_2e^{-i\omega_2t}+c.c.
\end{equation}
The frequency components of this field have arbitrary complex amplitudes
$E_{1,2}$ and elliptical polarizations described by unit vectors ${\bf
a}_{1,2}$. These vectors can be written in the spherical basis (${\bf e}_{\pm
1}=\mp({\bf e}_x\pm i{\bf e}_y)/\sqrt{2}$) as
\begin{eqnarray}\label{2}
&&{\bf a}_1=\sin(\varepsilon_1-\pi/4){\bf e}_{-1}+\cos(\varepsilon_1-\pi/4){\bf e}_{+1}\nonumber \\
&&{\bf a}_2=e^{-i\theta}\sin(\varepsilon_2-\pi/4){\bf e}_{-1}+ e^{i\theta}\cos(\varepsilon_2-\pi/4){\bf e}_{+1}\nonumber\\
&&-\pi/4\le\varepsilon_{1,2}\le\pi/4\,,
\end{eqnarray}
where $\varepsilon_{j}$ is the ellipticity parameter (angle) of the  $j$-th
component, $|\tan\varepsilon |$ is equal to the ratio of the minor semiaxes of
the ellipse  to the major one (see in Fig.1{\em a}), and the sign of
$\varepsilon$ governs the direction of the field vector rotation; $\theta$ is
the angle between the major semiaxes of the polarization ellipses. It is
assumed that atoms are being in a static magnetic filed ${\bf B}$ directed
along the $z$ axis (see in Fig.1{\em a}).

The field (\ref{1}) drives the two-photon resonance of $\Lambda$ type between
the two ground-state hyperfine levels with the total angular momenta $F_1$ and
$F_2$ (it does not matter integer or half-integer). The wavefunctions of the
Zeeman substates will be denoted as $|F_1,m\rangle$ and $|F_2,m\rangle$. In the
case of spectral resolution of the excited-state hyperfine structure the
two-photon resonance can be excited (by a suitable choice of the one-photon
detuning) via an isolated excited-state hyperfine level with the total angular
momentum $F_e$ (see in Fig.1{\em a}),with $\{ |F_e,\mu\rangle\}$ the
corresponding Zeeman substate wavefunctions. In what follows we shall use the
Dirac representation: $$e^{-i{\cal E}_{Fm}t/\hbar}|F,m\rangle \to|F,m\rangle
\,,$$ where ${\cal E}_{Fm}$ is the energy of the state $|F,m\rangle$ with
account for the Zeeman shift induced by the field ${\bf B}$. The dipole
interaction Hamiltonian $-(\widehat{\bf d}{\bf
E})=\widehat{V}+\widehat{V}^{\dagger}$ in the rotating wave approximation has
the form:
\begin{eqnarray}\label{V}
&&\widehat{V}=\\
&&d_{F_1F_e}E_1\sum_{m\mu q}e^{-i\delta^{(1)}_{\mu m}t}C^{F_e\mu}_{F_1m,1q}({\bf a}_1)^q |F_e,\mu\rangle \langle F_1,m|+\nonumber\\
&&d_{F_2F_e}E_2\sum_{m\mu q}e^{-i\delta^{(2)}_{\mu m}t}C^{F_e\mu}_{F_2m,1q}({\bf a}_2)^q |F_e,\mu\rangle \langle F_2,m|\,.\nonumber
\end{eqnarray}
Here $d_{F_1F_e}$ and $d_{F_2F_e}$ are the reduced dipole matrix elements of
the corresponding optical transitions $F_1\to F_e$ and $F_2\to F_e$ (see in
Fig.1{\em a}); $C^{F_em_e}_{F_jm_j,1q}$ denotes the Clebsch-Gordan coefficient;
$({\bf a}_j)^q$ are the contravariant spherical components of the polarization
vector ${\bf a}_j$  (see eq.(\ref{2})); and $\delta^{(j)}_{\mu
m}=\omega_j-({\cal E}_{F_e\mu} -{\cal E}_{F_jm})/\hbar$ (where $j=1,2$) are the
one-photon detunings.

The main objective of the present paper is to determine conditions of the
existence of dark states $|dark^{(m)}\rangle$, which are a coherent
superposition of the ground-state Zeeman sublevels $|F_1,m\rangle$ and
$|F_2,m\rangle$ with the same angular momentum projection $m$ with respect to
the $z$ axis:
\begin{equation}\label{dg}
|dark^{(m)}\rangle =A_1|F_1,m\rangle +A_2|F_2,m\rangle\,.
\end{equation}
These dark states nullify the interaction Hamiltonian (\ref{V}):
\begin{equation}\label{dark}
\widehat{V}\,|dark^{(m)}\rangle=0 \;
\end{equation}
at the exact two-photon $m$$-$$m$ resonance, when $\omega_1-\omega_2=({\cal
E}_{F_2m} -{\cal E}_{F_1m})/\hbar$. In the course of the optical pumping
process atoms are accumulated in the dark $m$$-$$m$ state, i.e. the pure
superposition state (\ref{dg},\ref{dark}) will be generated.

For the sake of simplicity we shall assume that the ground-state Zeeman
splitting exceeds significantly the width of the two-photon $m$$-$$m$
resonances, making them spectrally resolved. As is seen from Fig.1{\em b}, the
transitions between the states $|F_2,m\rangle$ and $|F_1,m\rangle$, induced by
the $\sigma_+$ or $\sigma_-$ circularly polarized filed components, form the
simple three-state $\Lambda$ systems. The dark states for each of those
$\Lambda$ systems can be written in the form:
\begin{eqnarray}\label{dark+}
&&|dark^{(m)}_+\rangle=N_{(+)}\times\\
&&\left[ |F_1,m\rangle -\frac{e^{-i\theta}E_1d_{F_1F_e}C^{F_e\,m+1}_{F_1m,11}\cos(\varepsilon_1-\pi/4)} {E_2d_{F_2F_e}C^{F_e\,m+1}_{F_2m,11}\cos(\varepsilon_2-\pi/4)}|F_2,m\rangle\right] \nonumber\\
\label{dark-}
&&|dark^{(m)}_-\rangle=N_{(-)}\times\\
&&\left[ |F_1,m\rangle -\frac{e^{i\theta}E_1d_{F_1F_e}C^{F_e\,m-1}_{F_1m,1\,-1}\sin(\varepsilon_1-\pi/4)} {E_2d_{F_2F_e}C^{F_e\,m-1}_{F_2m,1\,-1}\sin(\varepsilon_2-\pi/4)}|F_2,m\rangle\right]\nonumber,
\end{eqnarray}
where $N_{(\pm )}$ are the normalization constants. Obviously, the dark state
$|dark^{(m)}\rangle$, nullifying simultaneously the interactions with the
$\sigma_+$ and  $\sigma_-$ polarization components, is in existence if and only
if $|dark^{(m)}_+\rangle =|dark^{(m)}_-\rangle =|dark^{(m)}\rangle$. This
condition is satisfied, when:
\begin{eqnarray}\label{cond}
&&\frac{e^{-i\theta}C^{F_e\,m+1}_{F_1m,11}\cos(\varepsilon_1-\pi/4)} {C^{F_e\,m+1}_{F_2m,11}\cos(\varepsilon_2-\pi/4)}=\nonumber\\
&&=\frac{e^{i\theta}C^{F_e\,m-1}_{F_1m,1\,-1}\sin(\varepsilon_1-\pi/4)} {C^{F_e\,m-1}_{F_2m,1\,-1}\sin(\varepsilon_2-\pi/4)}\;.
\end{eqnarray}
It is more convenient to write down this equation as
\begin{equation}\label{cond1}
\frac{C^{F_e\,m+1}_{F_2m,11}C^{F_e\,m-1}_{F_1m,1\,-1}}
{C^{F_e\,m+1}_{F_1m,11}C^{F_e\,m-1}_{F_2m,1\,-1}} \;e^{i2\theta}=
\frac{\tan(\varepsilon_1+\pi/4)}{\tan(\varepsilon_2+\pi/4)}\;.
\end{equation}
Now we are ready to analyze in detail different variants of the total angular
momentum values $F_1$ and $F_2$.

\section{Alkali metal atoms}

Consider the case, when $F_1$ and $F_2$ differ by one, as it takes place, for
example, for alkali metal atoms. For definiteness, let $F_1=F$ and $F_2=F+1$.
Owing to the selection rules for dipole transitions the excited-state angular
momentum $F_e$ can acquire just two values $F$ and $F+1$.

Using explicit algebraic formula for the Clebsch-Gordan coefficients, we find
that the ratio in the l.h.s. of eq.(\ref{cond1}):
\begin{equation}\label{rel}
\frac{C^{F_e\,m+1}_{(F+1)m,11}C^{F_e\,m-1}_{Fm,1\,-1}}
{C^{F_e\,m+1}_{Fm,11}C^{F_e\,m-1}_{(F+1)m,1\,-1}}=-\frac{1+F-m}{1+F+m}
\end{equation}
does not depend on the value $F_e=F,F+1$, and it is negative. Therefore, taking
into account the positivity of the r.h.s. of eq.(\ref{cond1}), we obtain with
necessity the condition $e^{i2\theta}=-1$, i.e. the angle between the
polarization ellipse major semiaxes should be right ($\theta=\pi/2$). Thus, the
general universal bichromatic field configuration for the pure dark state
($|dark^{(m)}\rangle$) generation is $\varepsilon_1$$\perp$$\varepsilon_2$
configuration. The frequency component amplitudes $E_{1,2}$ can be arbitrary,
and the ellipticity parameters $\varepsilon_{1,2}$ obey the relationship:
\begin{equation}\label{e1-p-e2}
\frac{1+F-m}{1+F+m}=\frac{\tan(\varepsilon_1+\pi/4)}{\tan(\varepsilon_2+\pi/4)}\;.
\end{equation}

Let us distinguish the following particular configurations.\\ {\em A)} The
symmetric $\varepsilon$$\perp$$(-\varepsilon)$ configuration , where the
polarization ellipses are the same, but with opposite rotations of the field
vector. Inserting $\varepsilon_1=-\varepsilon_2=\varepsilon$ in
eq.(\ref{e1-p-e2}), we find the condition:
\begin{equation}\label{e-(-e)}
\sin(2\varepsilon)=-m/(1+F)\;,
\end{equation}
which couples the ellipticity parameter $\varepsilon$ with the  quantum numbers
$m$ and $F$.\\ {\em B)} The configuration {\em lin}$\perp$$\varepsilon$
($\varepsilon$$\perp${\em lin}), where one of the frequency component is
linearly polarized ($\varepsilon_1=0$ or $\varepsilon_2=0$). From
(\ref{e1-p-e2}) we obtain:
\begin{equation}\label{e-lin}
\tan\varepsilon=\pm m/(1+F)\;,
\end{equation}
where the sign ($+$)/($-$) corresponds to ({\em
lin}$\perp$$\varepsilon$)/($\varepsilon$$\perp${\em lin}), respectively.

Apart from these, the field configurations $\sigma_+$$-$$\sigma_+$ or
$\sigma_-$$-$$\sigma_-$ (where both frequency components have the same circular
polarization, i.e. $\varepsilon_1=\varepsilon_2= \pm\pi/4$) obey obviously the
condition (\ref{e1-p-e2}). However, these cases are degenerate, because in
addition to the superposition states $|dark^{(m)}\rangle$ there exists the trap
dark state (the end Zeeman state) $|F+1,m=\pm(F+1)\rangle$ insensitive to the
frequency difference $(\omega_1-\omega_2)$. As a result, one can not prepare a
pure superposition state and the atomic state will always correspond to a
statistical mixture of $|dark^{(m)}\rangle$ and $|F+1,m=\pm(F+1)\rangle$.

The two-photon  0$-$0 resonance is of great importance in view of possible
applications to atomic clocks. Let us consider this resonance separately.
Inserting  $m=0$ in eq.(\ref{e1-p-e2}), we see that
$\varepsilon_1=\varepsilon_2=\varepsilon$, i.e. the general configuration is
$\varepsilon$$\perp$$\varepsilon$, where the frequency components have the same
ellipticity and direction of rotation. Particular case of $\varepsilon =0$
corresponds to the {\em lin}$\perp${\em lin} field configuration, which has
been used in several works \cite{Taich,Jau,Zan} in order to increase the dark
resonance contrast. For instance, Zanon et al. \cite{Zan} used pulsed {\em
lin}$\perp${\em lin} field for the observation of the two-photon Ramsey-type
resonances. The push-pull optical pumping method developed in \cite{Jau} is
also equivalent to the {\em lin}$\perp${\em lin} configuration at the harmonic
amplitude modulation. It becomes clear, when the Fourier analysis is applied to
the problem. The dark resonance contrast increasing method described in our
papers \cite{Taich} is based on the constructive interference of the two-photon
transitions excited  by counter-propagating waves with orthigonal circular
polarizations ($\sigma_+$$-$$\sigma_-$ standing wave). This method can be also
interpreted in terms of the bichromatic {\em lin}$\perp${\em lin}
configuration, which is formed here locally in space, contrary to
\cite{Jau,Zan}.

It is interesting to note that the 0$-$0 dark state can be generated by a
multi-frequency field with arbitrary number of equidistant components. The
frequency difference has to coincide with the 0$-$0 transition frequency, the
amplitudes have to be the same as well as the phase differences between
adjacent harmonics. The frequency component polarizations are chosen in such a
way that any pair of adjacent components forms the
$\varepsilon$$\perp$$\varepsilon$ configuration. Under the above described
conditions the state $|dark^{(0)}\rangle$ is dark for any pair of adjacent
components as well as for the whole frequency comb. Such a frequency comb can
be formed by the push-pull optical pumping method \cite{Jau} with the pulsed
amplitude modulation.

From eq.(\ref{e1-p-e2}) it follows that the parameters $\varepsilon_{1,2}$ in
the $\varepsilon_1$$\perp$$\varepsilon_2$ configuration do not depend on the
excited state angular momentum $F_e$. Nevertheless, the explicit form of the
dark state $|dark^{(m)}\rangle$ can depend on $F_e$ in general case. As is seen
from eq.(\ref{dark+}), this dependence is governed by the ratio of dipole
moment matrix elements:
\begin{equation}\label{rel_d}
\frac{d_{F_1F_e}C^{F_e\,m+1}_{F_1m,11}}{d_{F_2F_e}C^{F_e\,m+1}_{F_2m,11}} \;.
\end{equation}
Such a dependence (if present) leads to spectacular consequences, if the
excited-state hyperfine structure is not spectrally resolved, as it usually is
in experiments in the presence of a buffer gas.

Using the well-known formula (see e.g.  \cite{varsh75}) for the reduced dipole
moment elements ($d_{F_1F_e}$ and $d_{F_2F_e}$), one can see that for the D2
line of alkali metal atoms the ratio (\ref{rel_d}) does depend on the
excited-state angular momentum $F_e$, and it equals to
\begin{equation}
-{\frac{(F-F_e+2)(F+F_e+3)\sqrt{1+F+m}}{(F-F_e-1)(F+F_e)\sqrt{1+F-m}}}\,.
\end{equation}
Consequently, the dark state $|dark^{(m)}\rangle$ will explicitly depend on
$F_e$. Due to this reason a common dark state, nullifying the interaction with
the field via all the excited-state hyperfine levels, is not in existence, as
it is illustrated by Fig.2{\em a} with the 0$-$0 resonance in D2 line of
$^{133}$Cs as an example. In Fig.2{\em a} one can clearly see that the ratio of
the dipole moment matrix elements for the  $\Lambda$ system formed via the
sublevel $|F_e=3,\mu =1\rangle$ differs from the similar ratio for the other
$\Lambda$ system formed through $|F_e=4,\mu =1\rangle$. Moreover, in the D2
line there are cycling transitions to the levels $F_e=F-1,F+2$ (see in
Fig.2{\em a}). According to the selection rules, these transitions can not form
the two-photon resonance, rather they destroy additionally the ground-state
coherence. Thus, in the absence of the spectral resolution in the excited state
the pure superposition state $|dark^{(m)}\rangle$ can not be realized in the D2
line.

A completely different type of situation occurs in the D1 line (see in
Fig.2{\em b}), where the ratio (\ref{rel_d}) does not depend on $F_e$:
\begin{equation}\label{rel_d1}
\frac{d_{FF_e}C^{F_e\,m+1}_{Fm,11}}{d_{(F+1)F_e}C^{F_e\,m+1}_{(F+1)m,11}}=
-\sqrt{\frac{1+F+m}{1+F-m}}\,.
\end{equation}
For example, the dark state $|dark^{(m)}\rangle$ in the symmetric
$\varepsilon$$\perp$$(-\varepsilon)$ configuration (\ref{e-(-e)}) for any $m$
has the form:
\begin{equation}\label{d_D1}
|dark^{(m)}\rangle=N\{ |F,m\rangle -i(E_1/E_2)|F+1,m\rangle\}\,.
\end{equation}
Note, the substate populations are controlled by the frequency component
amplitudes $E_1$ and $E_2$. Thus, in the D1 line the dark superposition states
$|dark^{(m)}\rangle$ can be prepared even in the absence of the spectral
resolution in the excited state. What is a key advantage of the D1 line (with
respect to the D2 line) in the dark resonance observation.

Besides, the dark state $|dark^{(0)}\rangle$ with $m=0$ in the D1 line can be
generated even if the ground-state hyperfine structure are not spectrally
resolved \cite{Jau}. To do this one has to use the above described
$\varepsilon$$\perp$$\varepsilon$ frequency comb with the total spectral width
much larger than the optical linewidth $\Gamma_{opt}$. For instance, in the
method of the paper \cite{Jau} it corresponds to the amplitude modulation by
pulses of duration $\tau\ll 1/\Gamma_{opt}$.

\section{Other possible variants}
For the sake of completeness, we give here the results for other possible
variants of the ground-state momentum values $F_1$ and $F_2$, which are allowed
by the selection rules for the two-photon resonance of $\Lambda$ type. These
results are follows from the analysis of eq.(\ref{cond1}). In practice such
variants can be realized, in principle, in several elements with complicated
ground-state structure, and also with the use of metastable levels.

The selection rules allow the following combination of the total angular
momenta: $F_1=F$, $F_2=F+2$ и $F_e=F+1$. In this case the dark state
$|dark^{(m)}\rangle$is generated by the $\varepsilon_1$$||$$\varepsilon_2$
field configuration, where the major ellipses semiaxes are parallel ($\theta
=0$), and the ellipticity parameters $\varepsilon_{1,2}$ obey the condition
\begin{equation}
\frac{(1+F-m)(2+F-m)}{(1+F+m)(2+F+m)}=\frac{\tan(\varepsilon_1+\pi/4)}{\tan(\varepsilon_2+\pi/4)}
\;.
\end{equation}
The amplitudes $E_{1,2}$ can be arbitrary.

The last variant is $F_1=F_2=F$, when the two-photon resonance can be excited
via the excited-state level with different momenta  $F_e=F,F\pm 1$. Here for
any $F_e$ the dark states are generated by the common (independent of $m$)
filed configuration $\varepsilon$$||$$\varepsilon$, where the frequency
components have the same elliptical polarization.

\section{Conclusion}
We have proposed the bichromatic field configurations, allowing to prepare
(with the use of the CPT effect) the pure superposition $m$$-$$m$ states for
arbitrary magnetic quantum number $m$. It has been shown that in the general
case of $m\ne 0$ it is necessary to use elliptically polarized fields. It has
been found that for alkali metal atoms in the absence of the excited-state
spectral resolution the superposition dark states can be only generated in the
D1 line. The obtained results have a wide spectrum of applications from atomic
clocks and magnetometers to quantum informatics. The preparation of the pure
$m$$-$$m$ states looks especially attractive in magnetometers, where it allows
one to get a high-contrast resonance with  maximal sensitivity to a magnetic
field.

Recently, we have got first experimental evidence
of the described theoretical predictions.
The corresponding results will published elsewhere.

We thank L.Hollberg, H.Robinson, J.Kitching, S.Knappe, and Y.-Y. Jau for
helpful discussions. This work is partially supported by RFBR
(grants 05-02-17086 and 04-02-16488) and by a grant INTAS-01-0855.

\begin{figure*}[h]\centerline{\scalebox{0.5 }{\includegraphics{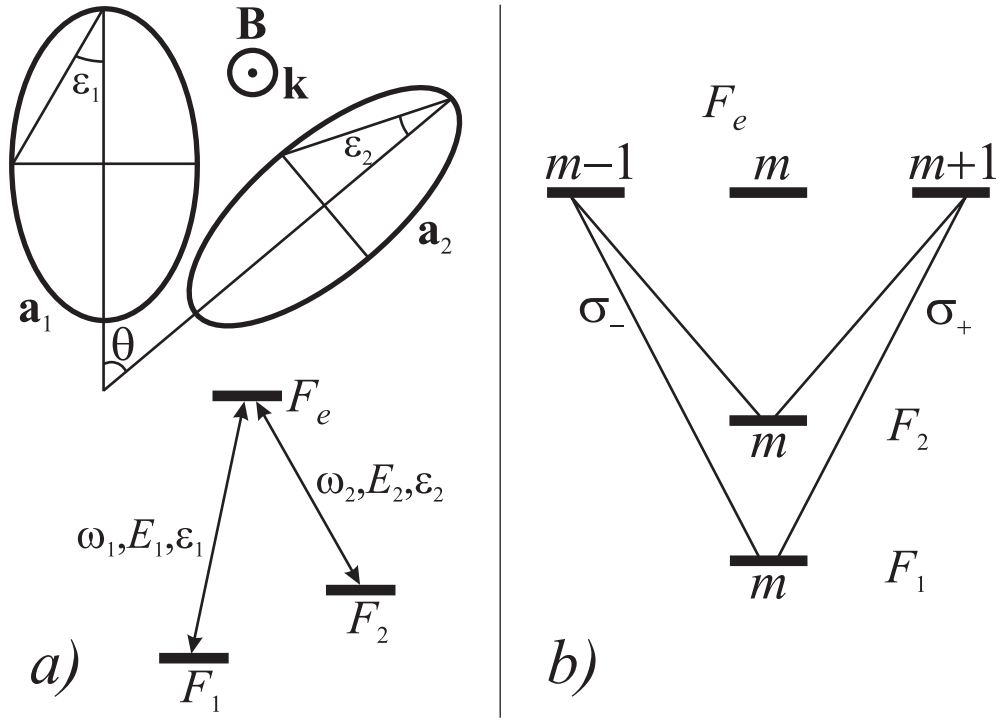}}}
\caption{Illustration to the problem statement: {\em a)} mutual orientation and
parameters of the polarization ellipses (${\bf a}_1,{\bf a}_2$), and the
general scheme of optical transitions, exciting the two-photon resonance of
$\Lambda$ type; {\em b)} scheme of the light-induced transitions driven  by the
$\sigma_+$ and $\sigma_-$ circularly polarized components for the $m$$-$$m$
resonance.} \label{fig1}
\end{figure*}

\begin{figure*}[h]\centerline{\scalebox{0.5}{\includegraphics{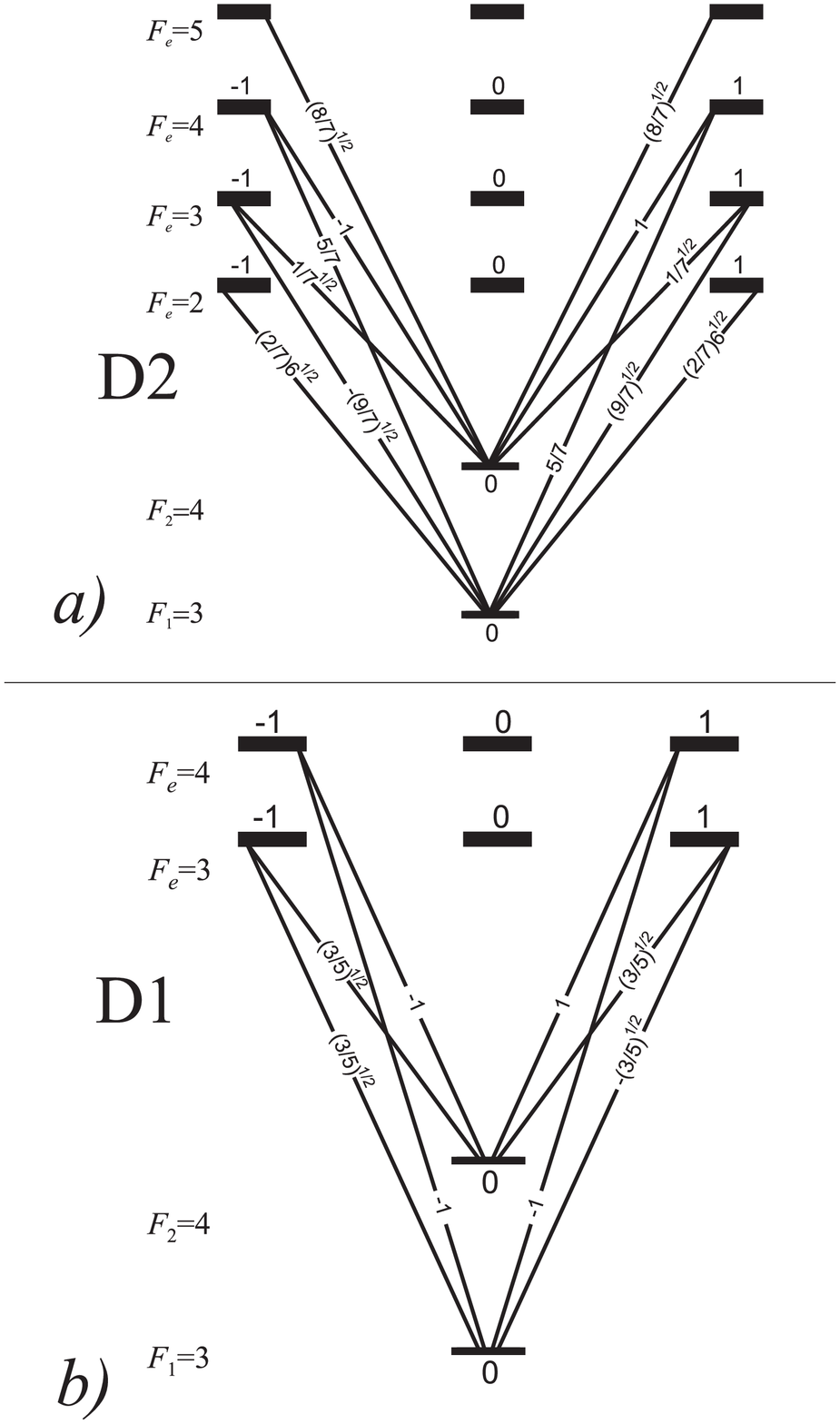}}}
\caption{Scheme of the light-induced transitions from the states
$|F_1,m=0\rangle$ and $|F_2,m=0\rangle$ in bichromatic filed. Numbers on lines
denote the relative values of the dipole moment matrix elements for $^{133}$Cs:
{\em a)} in the D2 line; {\em b)} in the D1 line.} \label{fig2}
\end{figure*}


\begin{thebibliography}{22}
\bibitem{Alz} G. Alzetta, A. Gozzini, L. Moi, and G. Orriols, Nuovo Cimento Soc. Ital. Fis., B {\bf 36}, 5 (1976).
\bibitem{Ar1} E. Arimondo and G. Orriols, Lett. Nuovo Cimento Soc. Ital. Fis. {\bf 17}, 33 (1976).
\bibitem{Gr} H.R. Gray, R.M. Whitley, and C.R. Stroud, Opt. Lett. {\bf 3}, 218 (1978).
\bibitem{St} M. Stahler, R. Wynands, S. Knappe, et al., Opt. Lett. {\bf 27}, 1472 (2002).
\bibitem{Van} J. Vanier, M.W. Levine, D. Janssen, et al., Phys. Rev. A {\bf 67}, 065801 (2003).
\bibitem{Taich} A.V. Taichenachev, V.I. Yudin, V.L. Velichansky et al., Pis'ma v ZhETF {\bf 80}, 265
(2004); S.V. Kargapoltsev, J. Kitching, V.L. Velichansky, et al., Laser Phys.
Lett., {\bf 1}, 495 (2004).
\bibitem{Jau} Y.-Y. Jau, E. Miron, A.B. Post, et al., Phys. Rev. Lett. {\bf 93}, 160802 (2004); Y.-Y.
Jau,  Ph.D. Thesis, Princeton University  (2004).
\bibitem{Zan} T. Zanon, S. Tremine, S. Guerandel, et al,. IEEE Trans. Instrum. Meas. {\bf 54}, (2005).
\bibitem{Hem} P.R. Hemmer, S. Ezekiel, C.C. Leiby, Opt. Lett. {\bf 8}, 440 (1983).
\bibitem{Kit} J. Kitching, S. Knappe, N. Vukicevic, et al, IEEE Trans. Instrum. Meas. {\bf 49}, 1313 (2000).
\bibitem{St2} M. Stahler, S. Knappe, C. Affolderbach, et al, Europhys. Lett. {\bf 54}, 323 (2001).
\bibitem{varsh75}  D.A. Varshalovich, A.N. Moskalev, V.K. Khersonsky,
{\em Quantum Theory of Angular Momentum} (World Scientific, Singapore, 1988).


\end{thebibliography}
\end{document}